# Kelvin probe characterization of buried graphitic microchannels in single-crystal diamond

E. Bernardi[1,2]*, A. Battiato[1,2], P. Olivero[1,2], F. Picollo[2,1], E. Vittone[1,2]

[1] *Physics Department, CNISM and "NIS" Inter-departmental centre, University of Torino, Torino, Italy*

[2] *Istituto Nazionale di Fisica Nucleare (INFN), Sezione di Torino, Torino, Italy*

\*      corresponding author: ettore.bernardi@unito.it

**Abstract**

In this work, we present an investigation by Kelvin Probe Microscopy (KPM) of buried graphitic microchannels fabricated in single-crystal diamond by direct MeV ion microbeam writing. Metal deposition of variable-thickness masks was adopted to implant channels with emerging endpoints and high temperature annealing was performed in order to induce the graphitization of the highly-damaged buried region. When an electrical current was flowing through the biased buried channel, the structure was clearly evidenced by KPM maps of the electrical potential of the surface region overlying the channel at increasing distances from the grounded electrode. The KPM profiling shows regions of opposite contrast located at different distances from the endpoints of the channel. This effect is attributed to the different electrical conduction properties of the surface and of the buried graphitic layer. The model adopted to interpret these KPM maps and profiles proved to be suitable for the electronic characterization of buried conductive channels, providing a non-invasive method to measure the local resistivity with a micrometer resolution. The results demonstrate the potential of the technique as a powerful diagnostic tool to monitor the functionality of all-carbon graphite/diamond devices to be fabricated by MeV ion beam lithography.

## I. Introduction

The existence at ambient conditions of different allotropic forms of carbon with radically different structural and electrical properties (most importantly, diamond and graphite) is an important feature allowing the fabrication of all-carbon devices for various technological applications. In this context, the employment of focused MeV ion beams in diamond is a versatile tool to create different structural forms of carbon by the progressive conversion of the $sp^3$-bonded diamond lattice to a $sp^2$-bonded amorphous/graphitic phase[1]. This effect is related to the ion-induced formation of structural defects (vacancies, interstitials and more extended complexes) in the crystal lattice, with a strongly non-uniform depth profile that determines the creation of highly-damaged layers at a depth of the order of micrometer within the crystal bulk, depending on the ion energy (see for example Fig. 1). Ion beam implantation in diamond was extensively applied for the fabrication of a broad range of devices: waveguides[2,3], photonic structures[4,5,6] micromechanical resonators[7,8]. In particular, the possibility of creating graphitic and electrically conductive regions allowed the fabrication of infrared radiation emitters[9], field emitters[10], bolometers[11], biosensors[12] and ionizing radiation detectors[13, 14].

The charge conduction mechanisms in amorphized/graphitized diamond have been investigated in previous works[15,16,17]. In the case of buried conducting layers and channels, the analyses were based on current-voltages characterizations carried out by probing 2 or 4 terminals emerging at the surface through conductive columns fabricated by laser induced graphitization[18], high-voltage-induced thermal breakdown[19] or by modulating ion penetration depths by multiple energy ion implantation[20] or variable-thickness metallic masks[21].

However, to our best knowledge, because of the inaccessibility of the buried conductive regions, no local microscopic analyses of their electrical properties have been so far performed. They could be effectively explored by scanning probe microscopy techniques such as Electric Force Microscopy (EFM)[22] and Kelvin Probe Microscope (KPM)[23,24], but these techniques are mainly used to investigate the surface electrical properties of semiconductors and insulators, such as the potentials of different terminations of diamond surface[25,26]. It has to be underlined that KPM has been used for the study of the buried interface between diamond and a thin film (25 nm) of polypyrolle[27] and , recently, for the spectrally and spatially resolved photovoltage measurement of a Schottky junction made of a 4 nm thin tungsten-carbide layer on oxygen-terminated boron-doped diamond [28].On the other hand, they have been rarely used to study deeper buried structures, as in the case of the detection of metal-polymer interfaces buried under a 100 nm polymer thick film[29].

In this work we demonstrate the possibility of imaging the electronic properties of deep (1 μm) buried graphitic structures in diamond by scanning probe microscopy. KPM maps and profiles were taken with sub-micrometric spatial resolution in the presence of a steady DC current flowing in the channels, in order to characterize the device in working conditions

Results are analysed by modelling both the capacitance couplings and the voltage drops between the buried channels and the conductive surface.

## II. Experimental

**A. Ion Beam Fabrication**

Ion implantation was performed on a synthetic single crystal diamond produced by HPHT (High Pressure High Temperature) process and classified as type Ib on the basis of single substitutional nitrogen ([N] ~ 10 ÷ 100 ppm) content. The crystal was cut along the [100] crystallographic direction and it was polished on the large face that was microfabricated and characterized. The sample was implanted at the AN2000 accelerator of the INFN - Legnaro National Laboratories[30] with a scanning focused 1.2 MeV He$^+$ ion beam at a fluence of ~$10^{17}$ cm$^{-2}$. The beam spot size was ~10 µm and the beam current was comprised between 2 nA and 3 nA. Before the implantation process, the diamond surface was covered with a 1 µm thick copper layer in order to reduce the ion penetration depth to the desired value (1 µm). Furthermore, variable thickness masks were employed in order to realize damaged structures with endpoints emerging up to the surface[21]. Figure 1 shows the depth profile of the ion-induced structural damage, being parameterized as the volumetric density of induced vacancies, as resulting from SRIM 2013.00 Monte Carlo code[31] in "Detailed calculation with full damage cascades" mode, by setting a displacement energy value of 50 eV[32]. As already pointed out in previous works[33], the vacancy-density profile was obtained by modelling the effect of cumulative ion implantation on the number of vacancies with a simple linear approximation, which does not take into account complex processes such as self-annealing, ballistic annealing and defect-defect interaction, therefore such value can only be considered as an effective parameter quantifying the induced damage density. The implantation resulted in the formation of sub-superficial amorphized layers, structured in one longitudinal channel and two transverse channels, located at ~1 µm below the surface.

After removal of the copper mask, thermal annealing at 1000 °C for 1 hour was performed to induce the graphitization of the highly-damaged buried region, where the vacancy densities exceeds a threshold value estimated as ~$9\times10^{22}$ cm$^{-3}$ in previous works[34,21]. As a result, the buried amorphized regions converted to ~200 nm thick graphitic channels located at the above-mentioned depth (see the highlighted region of the profile in figure 1). Finally, the endpoints were contacted with metal pads: a Cr/Cu deposition was followed by 400 °C heating for 1 hour to create a conductive carbide compound with chromium. In figure 2 an optical micrograph (a) and a 3D model of the final device (b) are shown: two transversal channels emerge at the diamond surface in four locations highlighted by red circles. At each side of the main longitudinal channel, the pair of emerging points is short-circuited through the common surface metal deposition.

**B. Current-voltage characteristics**

The current-voltage characteristics of the channels were measured at room temperature with a Keithley 2636 electrometer in a standard two-electrode configuration, by using two micro-tip probes in contact with the two Cr/Cu metal electrodes deposited onto the channel endpoints. Surface conductivity was measured using two similar control electrodes not connected to the graphitic channel.

**C Atomic force microscopy**

A Park Scientific XE-100 AFM equipped with rectangular Au-Cr-coated cantilevers (NSC-14 Cr-Au, MikroMash, spring constant of ~5.7 N m$^{-1}$ and resonance frequency between 160 kHz and 170 kHz) was used to perform a morphological characterization of the microfabricated sample. The Typical scanning areas were 35×35 µm$^2$ and the scanning rate was 0.5 Hz. The oscillation frequency of the cantilever was near the resonance on the upper-side. The morphology of the sample was recorded in non-contact, amplitude modulation mode.

**D Kelvin Probe Microscopy**

In KPM measurements the AC voltage applied to the cantilever tip was set to $V_{ac} = 2.5$ V and a DC voltage $V_{app} = 3.5$ V was applied to one of the surface electrodes while the other (left) electrode was grounded in order to have a current flowing through the longitudinal buried channel (figure 2b). The electrostatic forces at frequency $v = 17$ kHz were minimized by a feed-back loop controlling the DC voltage $V_{dc}$ applied to the tip. Typical scanning areas were 35×5 µm$^2$ at a scanning rate of 0.2 Hz. In all the experiments, $V_{ac}$ and $V_{dc}$ were applied directly by the control unit of the AFM microscope, whereas a Keithley 614 electrometer was used as external voltage source for $V_{app}$. The cantilever was oriented parallel to the long axis of the electrodes to minimize artifacts due to the capacitance between the cantilever beam and the electrodes[35].

## III. Results and discussion

Voltage-current characteristics of the surface and of the buried channels (data not reported here) exhibit a linear ohmic behaviour, in agreement with was reported in Picollo et al.[21]. The electrical resistance of the buried graphitic channel results $R_c$ =3.44 kΩ, a value which is negligible compared to the resistance of the surface ($R_s$ = 493 GΩ), which is attributed to a surface conductive layer induced by the high temperature annealing of the sample[36].

The topography profile above the buried channel is reported in figure 2c, which shows a pronounced surface swelling of ~100 nm localized at the implanted area. The presence of surface swelling in implanted diamonds is a well know effect[37,38,39] due to the lower density of the graphitic channel with respect to the one of the surrounding diamond crystal.

As shown in figures 3a and 3b, KPM maps of dimensions 35×5 μm$^2$ were acquired across the longitudinal channel at different distances from the left electrode (points "a" and "b" in figure 2a).

The KPM maps clearly show a central plateau in correspondence of the buried channel. As expected, the KPM maps are not remarkably influenced by the topography of the surface (i.e. swelling shown in the AFM map in figure 2c) as proven by the inversion of the contrast, when probing regions close to the grounded (figure 3a) or the biased (figure 3b) electrode.

To analyse in more detail the behaviour of the contrast in the KPM maps, transversal profiles (i.e. along the *y* axis) were extracted from KPM maps similar to those shown in figures 3a and 3b and centred in different positions along the channel (coloured dashed segments in figure 2a). Figure 3c shows these profiles relevant to 7 different positions distributed along the *x* axis from the grounded (left) to the biased (right) electrode. Figure 4 summarizes the different behaviour of the average KPM signals relevant to the surface (i.e. far from the buried channel, $^{KPM}V_S$) and to the buried channel ($^{KPM}V_C$) as a function of the *x* position. In order to remove the effects of the contact potential and of the intrinsic surface potential[40], all the KPM signals have been subtracted to the reference signal measured with both grounded electrodes. Moreover, this procedure minimizes any possible influence of surface contaminations on the KPM measurements[41,42].

From the inspection of figure 4, the following considerations can be made:

i) the KPM signals from both the regions directly above and aside the buried channel increase when moving from the grounded electrode to the electrode biased at $V_{app} = +3.5$ V;

ii) near the grounded left electrode, the KPM signal on the region directly above the buried channel is larger than the baseline, while the opposite happens near the right electrode; in between, a cross-over between the two profiles is observed.

To understand this behaviour we have to consider that the KPM signal corresponds to the $V_{dc}$ that minimize local electrostatic force $F_\omega$ induced by several conductive elements present in the system[43,44]. In this case the local force sensed by the "tip + cantilever" system at a distance $x$ from the left electrode $F(x)$ is the sum of the contributions relevant to all the conductive elements, namely the surface conductive layer ($F_0$), the longitudinal buried channel ($F_1$), the left and right electrodes ($F_2$, $F_3$) and the left and right transversal channel ($F_4$, $F_5$). The $i$-th force $F_i(x)$ at point $x$ depends on the potential ($V_i(x)$) generated by the $i$-th conductive element and on the capacitive coupling $K_i(x)$ between the "tip + cantilever" system and the $i$-th conductive element, i.e.:

$$F(x) = \sum_i F_i(x) = \sum_i K_i(x)[V_{dc} - V_i(x)] \quad (1)$$

The KPM signal ($^{KPM}V$) is then given by:

$$^{KPM}V = \frac{\sum_i K_i(x) V_i(x)}{\sum_i K_i(x)} \quad (2)$$

Equation (2) implies that the KPM signal is the weighted average of the potentials associated to the different conductive elements, the weighting factors being the capacitive coupling of the conductive elements with the tip. We note that these terms decrease with the increasing of the distance from the conductive element to the "tip + cantilever" system[43], hence, far from the electrodes, the dominant terms in the sums of equation 4 are related to the surface ($K_0$) and to the buried channel ($K_1$).

Regarding the contribution in equation (2) relevant to the surface ($V_0(x)$) it can expressed as the interelectrode (w = 220 μm) potential drop:

$$\qquad - \qquad \qquad (3)$$

$V_0(x)$ is indicated by the dashed line in figure 4.

Furthermore, we can safely assume that far away from the buried channel, the term related to the longitudinal buried channel in equation (2) ($i = 1$) can be neglected, and, by plotting $V_0$ as function of the experimental KPM signal relevant to the surface ($^{KPM}V_S$), it is apparent that the contribution of all the other terms provide a linear relationship (figure 5a), i.e. $\qquad$. The coefficients $a = (0.65 \pm 0.06)$ V and $b = (0.62 \pm 0.03)$, extracted from the linear fit of the experimental data, can be considered as defining the transfer function, which correlate the KPM signal to the real potential at point $x$. In particular, at point $x$ located onto the buried channel, the KPM signal $\qquad$, filtered from the contribution of the left and right electrodes ($F_2$, $F_3$) and transversal channel ($F_4$, $F_5$) can be evaluated through the expression:

$$\qquad \qquad (4)$$

as shown in figure 5b.

Actually, the macroscopic electrostatic tip-sample interaction can be calculated from the potential energy:

$$\qquad - \qquad \qquad - \qquad \qquad (5)$$

stored in the two capacitors in series formed between the tip/surface and surface/channel. The free energy ( ) of the system can then be calculated by the sum of the two capacitor contributions and by the work done by the sources[45]:

$$\qquad ( \qquad ) $$
$$\overline{\qquad} \qquad \overline{\qquad} \qquad (6)$$

By virtue of the high resistance of the surface, in equation (5) the work done by the voltage source was consider acting only in maintaining fixed $V_1$, whereas $V_0$ is let free to change according to the potential related to the charge induced by $V_1$ onto the surface, minimizing the electrostatic energy along the z direction, it is found that $^{KPM}V_C$ is equal to $V_1$.

Such an interpretation is further corroborated by the potential drop calculated from elementary circuital analysis of the five resistances drawn in the inset in figure 5b, i.e. four parallel transversal resistances of lengths $d_i$ ($i = 1, 4$) and one longitudinal resistance of length $d_L+d_R+w$. Assuming a constant resistivity of the graphitic channel and a constant section of the buried channel, the potential $V_1(x)$ is given by:

$$V_1(x) = \frac{(\ )(\ )}{(\ )+(\ )} \quad (7)$$

it is worth noting that, apart from a slight shift, $V_1$ is close to the experimental data corrected by the linear transfer function ( 　　　 , see figure 5b.

Finally, it is worth remarking that the abovementioned analysis provides a method to locally evaluate the channel resistivity. In fact, the slope of the experimental potential drop along the channel (─── ) is connected to the graphitic channel resistivity $\rho$ by the following relationship:

$$\frac{(\ \ )}{-\ \ \ \ \ } \quad (8)$$

where $A$ is the channel section (200 nm × 12 μm), $I$ is the current (1.04 mA) flowing through the graphitic channel. Such a value is in good agreement with the typical resistivity values measured for standard polycrystalline graphite ($\rho = 3.5$ mΩ cm)[21,46].

It is worth remarking that the linear behaviour of the 　　　 data point implies the homogeneity of the resistance along the main axes of the channel.

Lastly, it has to be underlined that the asymmetries in the device geometry do not significantly affect our key findings. Indeed, the lack of symmetry in the positioning of the metal electrodes results only in a shift of the cross-over between the two profiles of figure 4 from the mid-point between the two electrodes to a position slightly closer to the left electrode. Furthermore, the definition of two transversal channels and their slight asymmetry with respect to the axis of the longitudinal channel result substantially in the addition of two resistors in parallel at each side of the buried channel equivalent circuit depicted in the inset of figure 5b.

## IV. Conclusions

Sub-superficial graphitic microchannels fabricated in single-crystal diamond with a deep ion beam fabrication technique were characterized by mapping the electrostatic properties of the sample by probing in non-contact mode the electrostatic tip-sample force. KPM was proven to be suitable to map graphitic channels buried at a depth of 1 $\mu$m. KPM maps represent the actual electrical images of the graphitic channels, since they are practically insensitive from the surface morphology. The model adopted to interpret these maps provides a non-invasive method to measure the local resistivity of buried conductive channels. For the case of study, the constant potential drop across the buried channel evidences a uniform distribution of resistivity, whose value is in agreement with that of polycrystalline graphite.

These results have significant implications for the fabrication of all-carbon graphite/diamond devices; in facts, they have demonstrated the potential of the technique to non-invasively map with a micrometer resolution and in working conditions, the electrical properties of more complex buried graphitic patterns, which can be directly written by MeV ion beam lithography in diamond[47].

## Acknowledgments

This research activity was supported by the following projects: FIRB project D11J11000450001 funded by MIUR; "A.Di.NTech." project D15E13000130003 and "2011-Linea 1A" project ORTO11RRT5 funded by the University of Torino-Compagnia di San Paolo; "DiNaMo" (young researcher grant, project n° 15766) by National Institute of Nuclear Physics.

## Figures Captions

**Fig. 1** Vacancy density profile induced by 1.2 MeV He$^+$ ions at a fluence of $1\times10^{17}$ cm$^{-2}$ in diamond covered by a 1-µm-thick copper layer as calculated by SRIM2013.00 Monte Carlo simulations. The dashed region in the graph represents the highly damaged buried region where the thermal annealing induces the graphitization, due to the vacancy densities exceeding a critical threshold.

**Fig. 2 (a)** Optical micrograph of the buried channels, which lie at ~1 µm below the sample surface. The emerging end-points are highlighted by red circles. All channels are 12 µm wide and the inter-electrode distance is 220 µm. White letters indicates the locations of the KPM maps in figures 3 (a,b): dashed lines indicates the locations of the KPM profiles shown in figure 3(c). **(b)** Schematics (not to scale) of the sample and of the KPM experimental setup. Both DC ($V_{dc}$) and AC ($V_{ac}$) voltages are applied to the conductive AFM microtip. A second independent DC voltage ($V_{app}$) is applied across the buried graphitic channel. **(c)** Topography profile acquired across the buried channel. The horizontal axis is orthogonal to the graphitic channel.

**Fig. 3** KPM maps collected from surface regions above the buried channel lying closer to the left (**a**) and the right (**b**) electrode (regions "a" and "b" in figure 2 (a)). The left electrode is grounded; the right electrode is biased at $V_{app}$ = +3.5 V. Maps were acquired with the fast scanning direction orthogonal to the channel axis. **(c)** KPM profiles collected across the buried channel at different distances, indicated by dashed lines in figure 2 (a), from the left electrode.

**Fig. 4** Plots of the KPM signal as a function of the distance from the left electrode, both for regions directly above (black square dots) and away (red circular dots) from the buried channel. The black dashed line represents $V_0(x)$ given by eq. (2).

**Fig. 5 (a)** Predicted potential at the surface $V_0$, defined by eq. 3, as function of experimental potential KPM signal $^{KPM}V$. The dot-dashed line is the best-fit line. **(b)** Experimental ($^{KPM}V_c(x)$, red filled circles) and corrected by the linear transfer function in eq. (4) ($^{KPM}V_c^{corr}(x)$, red hollow circles) potential drop from KPM signal. The line is the predicted potential drop (eq. 7) derived from the equivalent circuit schematically drawn in the inset.

**Fig. 1**

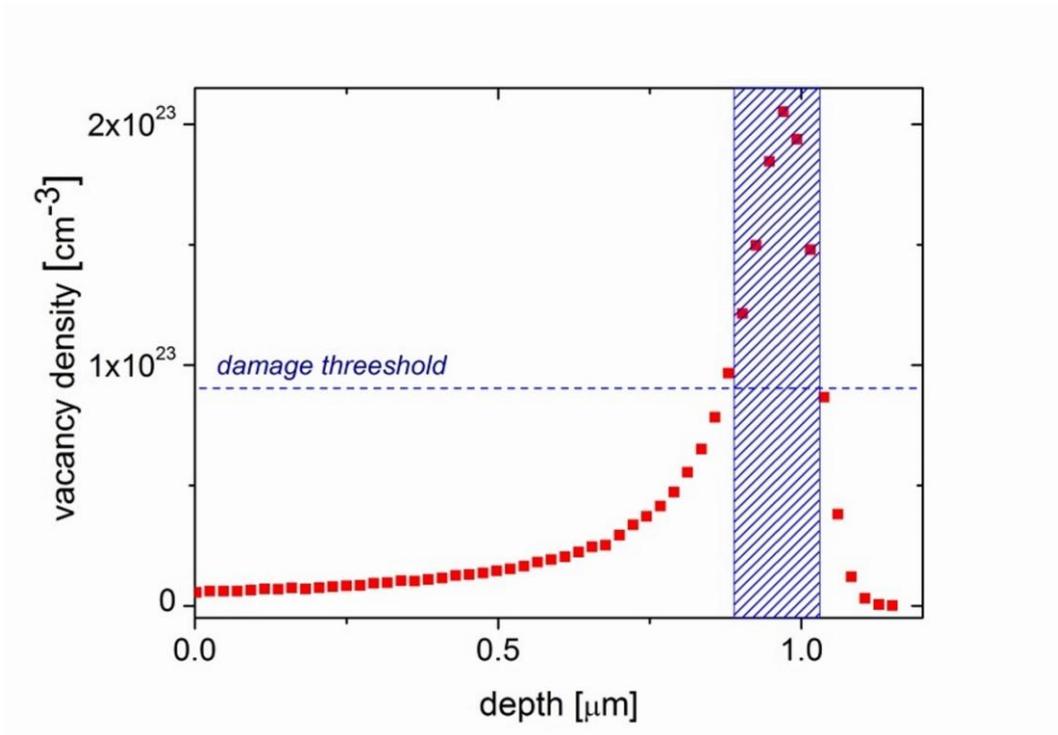

**Fig.2**

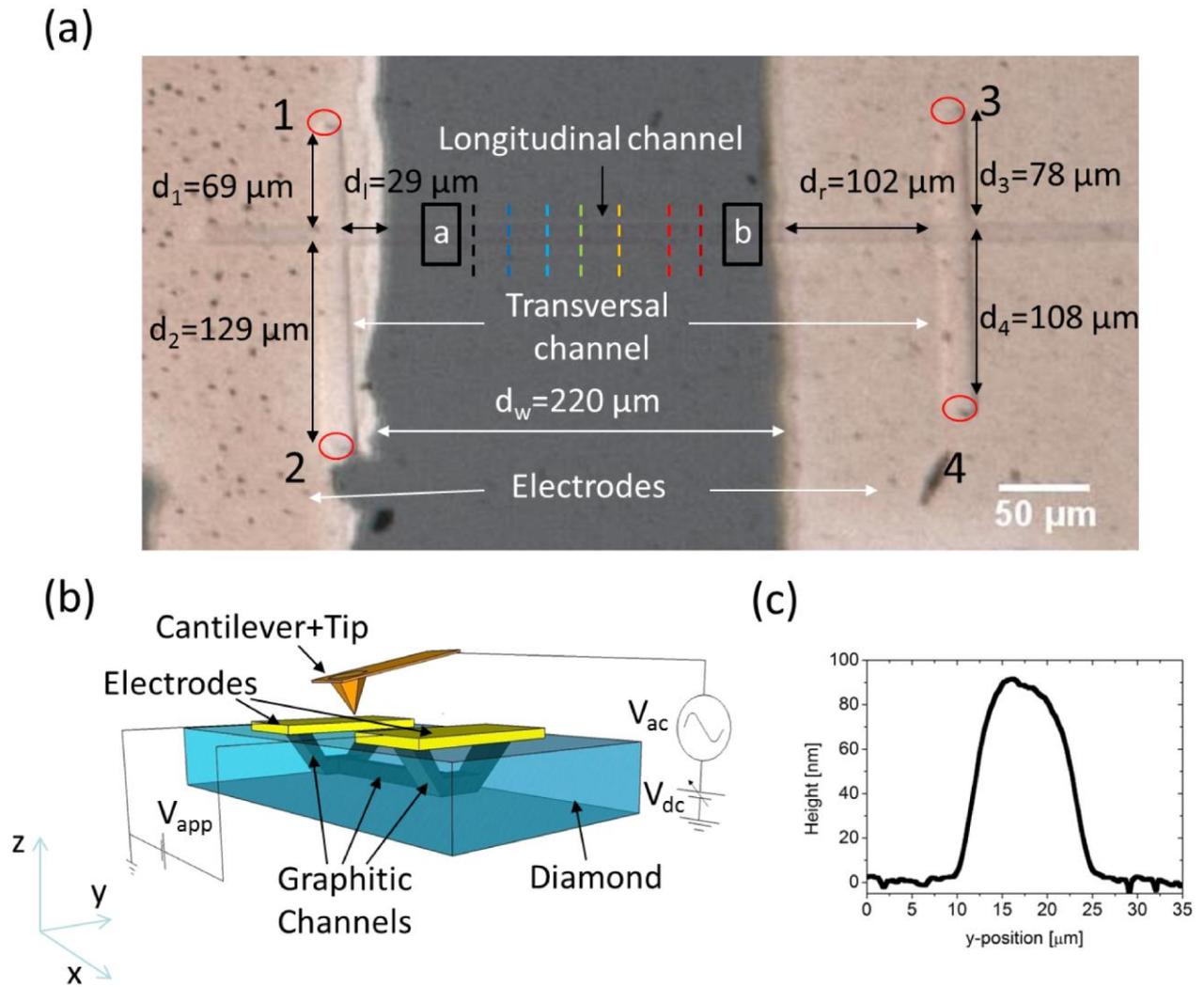

**Fig. 3**

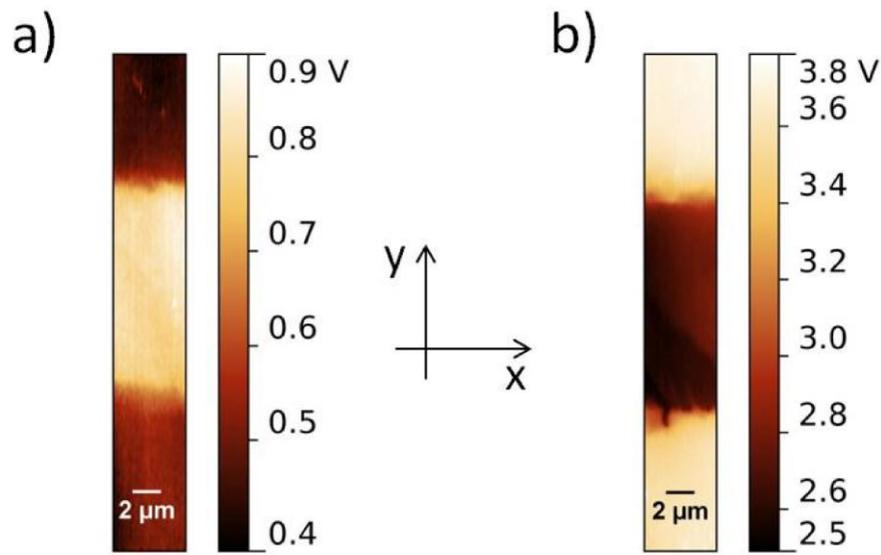

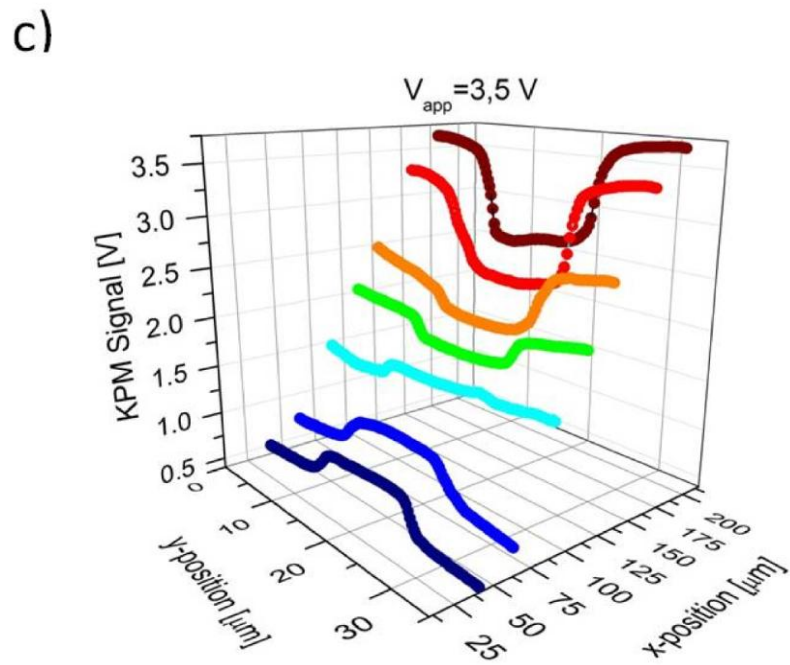

**Fig. 4**

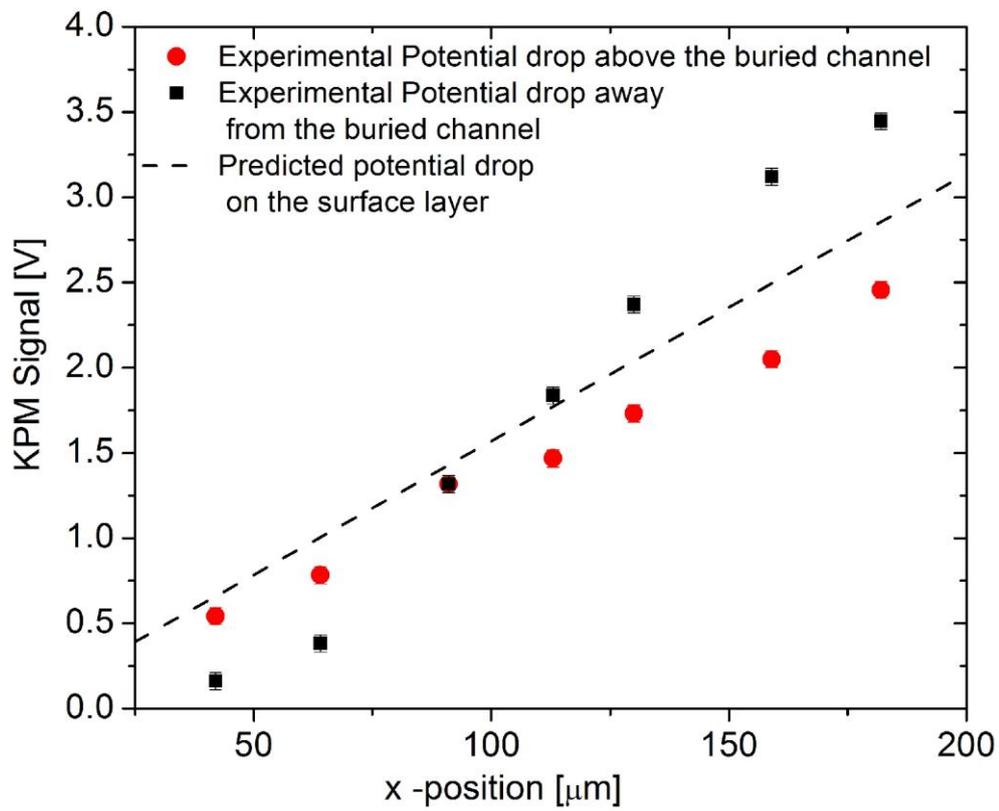

**Fig. 5**

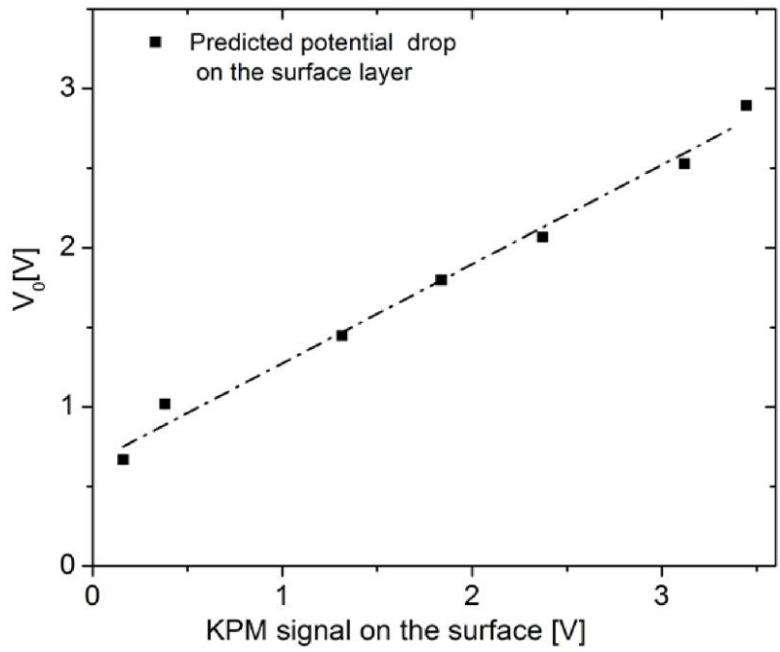

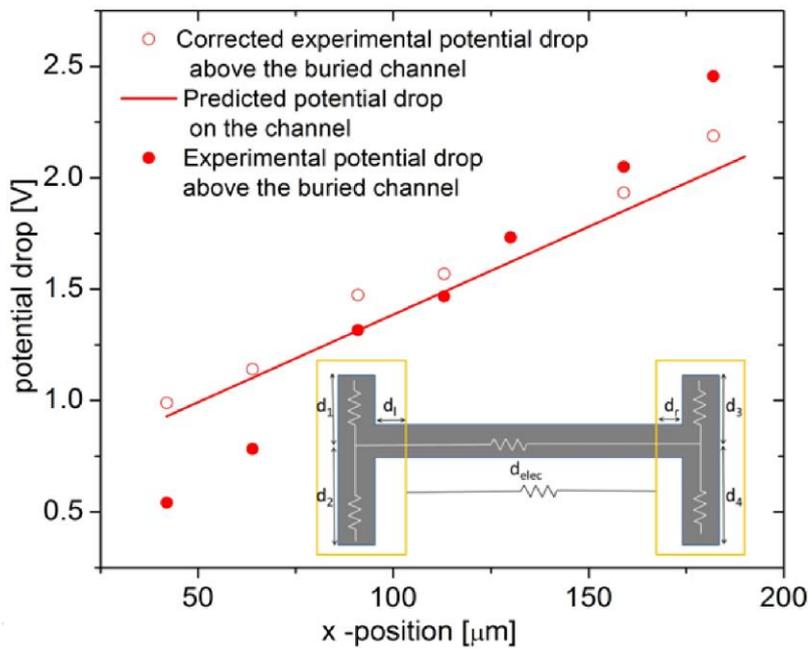

References

[1] S. Prawer, D. N. Jamieson, R. Kalish, Physical Review Letters **69**, 2991 (1992)

[2] P. Olivero, S. Rubanov, P. Reichart, B. C. Gibson, S. T. Huntington, J. Rabeau, A. D. Greentree, J. Salzman, D. Moore, D. N. Jamieson, S. Prawer, Advanced Materials **17**, 2427 (2005)

[3] M. P. Hiscocks, K. Ganesan, B. C. Gibson, S. T. Huntington, F. Ladouceur, S. Prawer, Optics Express **16**, 19512 (2008)

[4] I. Bayn, B. Meyler, A. Lahav, J. Salzman, R. Kalish, B. A. Fairchild, S. Prawer, M. Barth, O. Benson, T. Wolf, P. Siyushev, F. Jelezko, J. Wrachtrup, Diamond and Related Materials **20**, 937 (2011)

[5] J. C. Lee, I. Aharonovich, A. P. Magyar, F. Rol, E. L. Hu, Optics Express **20**, 8891 (2012)

[6] B. R. Patton, P. R. Dolan, F. Grazioso, M. B. Wincott, J. M. Smith, M. L. Markham, D. J. Twitchen, Y. Zhang, E. Gu, M. D. Dawson, B. A. Fairchild, A. D. Greentree, S. Prawer, Diamond and Related Materials **21**, 16 (2012)

[7] M. Liao, S. Hishita, E. Watanabe, S. Koizumi, Y. Koide, Advanced Materials **22**, 5393 (2010)

[8] M. K. Zalalutdinov, M. P. Ray, D. M. Photiadis, J. T. Robinson, J. W. Baldwin, J. E. Butler, T. I. Feygelson, B. B. Pate, B. H. Houston, Nano Letters **11**, 4304 (2011)

[9] S. Prawer, A. D. Devir, L. S. Balfour, R. Kalish, Applied Optics **34**, 636 (1995)

[10] A. V. Karabutov, V. G. Ralchenko, I. I. Vlasov, R. A. Khmelnitsky, M. A. Negodaev, V. P. Varnin, I. G. Teremetskaya, Diamond and Related Materials **10**, 2178 (2001)

[11] A. I. Sharkov, T. I. Galkina, A. Y. Klokov, R. A. Khmelnitskii, V. A. Dravin, A. A. Gippius, Vacuum **68**, 263 (2003)

[12] F. Picollo, S. Gosso, E. Vittone, A. Pasquarelli, E. Carbone, P. Olivero, V. Carabelli, Adv. Mater. **25**, 4696 (2013)

[13] P. J. Sellin, A. Galbiati, Applied Physics Letters **87**, 093502 (2005)

[14] J. Forneris, V. Grilj, M. Jakšić, A. Lo Giudice, P. Olivero, F. Picollo, N. Skukan, C. Verona, G. Verona-Rinati, E. Vittone, Nucl. Instr. Meth. in Phys. Res. B, **306**, 181-185 (2013)

[15] J. F. Prins, Physical Review B **31**, 2472-2478 (1985).

[16] R. Kalish, A. Reznik, K. W. Nugent, S. Prawer, Nucl. Instr. Meth. in Phys. Res. B **148**, 626 (1999)

[17] E. Baskin, A. Reznik, D. Saada, J. Adler, R. Kalish, Physical Review B **64**, 224110 (2001)

[18] R . Walker, S. Prawer, D.N. Jamieson and K.W. Nugent, Appl. Phys. Lett. **71**, 1492 (1997)

[19] A.A. Gippius, R.A. Khmelnitskiy, V.A. Dravin and S.D. Tkachenko, Diamond Relat. Mater. **8**, 1631 (1999)

[20] T .I. Galkina, A.Y. Klokov, A.I. Sharkov, R.A. Khmelnitski, A.A. Gippius, V.A. Dravin, V.G. Ral'chenko and A.V. Savel'ev, Phys. Solid State **49** 654 (2007)

[21] F. Picollo , D. Gatto Monticone , P. Olivero , B. A. Fairchild , S. Rubanov , S. Prawer , E. Vittone, New Journal of Physics **14**, 053011 (2012)

[22] P. Girard, Nanotechnology **12**, 485–490 (2001)

[23] Y. Martin, D.W. Abraham and H.K. Wickramasinghe, Appl. Whys. Len., **52,** 1103 (1988).

[24] W. Melitz, J.Shen, A. C. Kummela, S. Lee, Surface Science Reports **66**, 1 (2011)

[25] B. Rezek, C.E. Nebel, Diamond and Related Materials **14,** 466 (2005)

[26] M. Tachiki,. Y. Sumikawa, M. Shigeno., H. Kanazawa, T. Banno, K. Soup Song, H. Umezawa, H. Kawarada, Surface Science **581,** 207 (2005)

[27] B. Rezek , J. Čermák, A. Kromka, M. Ledinský, J. Kočka, Diamond Relat. Mater. **18**, 249 (2009)

[28] J. Čermák, Y. Koide, D. Takeuchi and B. Rezek, Journal of Applied Physics **115**, 053105 (2014)

[29] M.M. Marzec, K. Awsiuk, A. Bernasik, J. Rysz , J. Haberko ,W. Łużny , A. Budkowski, Thin Solid Films **531,** 271(2013)

[30] D. Bollini, F. Cervellera, G.P. Egeni, P. Mazzoldi, G. Moschini, P. Rossi, V. Rudello, Nucl. Instr. and Meth. A **328**, 173 (1993)

[31] [http://www.srim.org/]

[32] W. Wu, S. Fahy, Physical Review B **49**, 3030 (1994)

[33] F. Bosia, N. Argiolas, M. Bazzan B. Fairchild, D. Greentree, D. W. M. Lau, P. Olivero, F. Picollo, S. Rubanov, S. Prawer, J. Phys. Condens. Matter 25, 385403 (2013)

[34] P. Olivero, S. Rubanov, P. Reichart, B. C. Gibson, S. T. Huntington, J. R. Rabeau, A. D. Greentree, J. Salzman, D. Moore, D. N. Jamieson, S. Prawer, Diamond and Related Materials **15**, 1614-1621 (2006)

[35.] K. Puntambekar, P. Pesavento, and C. Frisbie, Applied. Physics. Letters **83,** 5539

[36] Yu. V. Butenko, V. L. Kuznetsov, A. L. Chuvilin, V. N. Kolomiichuk, S. V. Stankus, R. A. Khairulin, and B. Segall, Journal of Applied Physics **88**, 4380 (2000)

[37] J. F. Prins, T. E. Derry, J. P. F. Sellschop, Physical Review B **34**, 8870-8874 (1986)

[38] F. Bosia, S. Calusi, L. Giuntini, S. Lagomarsino, A. Lo Giudice, M. Massi, P. Olivero, F. Picollo, S. Sciortino, A. Sordini, M Vannoni, E Vittone. Nucl. Instr. Meth. B **268,** 2991, (2010)

[39] M. Piccardo, F. Bosia, P. Olivero, N. Pugno, Diamond & Related Materials **48**, 73 (2014)

[40] D. J. Bayerl and X. Wang Adv. Funct. Mater. **22**, 652 (2012)

[41] H. Bluhm, T. Inoue, M. Salmeron, Surface Science **462**, 602 (2000)

[42] W.W.R. Araoujo, M.C. Salvadori, F.S. Teixeira, M. Cattani, I.G. Brown, Microscopy Research And Technique **75**, 977 (2012)

[43] H.O. Jacobs, P. Leuchtmann, O.J. Homan and A. Stemmer, Journal of Applied Physics **84**, 1168 (1998)

[44] D. Brunel, D. Deresmes, and T.Mélin, Applied Physics Letters **94**, 223508 (2009)

[45] A. Sadeghi, A. Baratoff, S. Alizera Ghasemi, S. Goedecker, T. Glatzel, S Kawai, and E. Meyer Physical Review B **86**, 075407 (2012)

[46] J.D. Cutnell and K.W Johnson, Resistivity of Various Materials in Physics (New York: Wiley) 2004

[47] F. Picollo, A. Battiato, E. Bernardi, L. Boarino, E. Enrico, J. Forneris, D. Gatto Monticone, P. Olivero, arXiv:1412.2212